# Revitalizing Endangered Languages: AI-powered language learning as a catalyst for language appreciation

Dinesh Kumar Nanduri
College of Information Studies, University of Maryland - College Park, USA, dnanduri@umd.edu

Elizabeth M. Bonsignore
College of Information Studies, University of Maryland, College Park, USA, ebonsign@umd.edu

**ABSTRACT**

According to UNESCO, there are nearly 7,000 languages spoken worldwide [1], of which around 3,000 languages are in danger of disappearing before the end of the century [2]. With roughly 230 languages having already become extinct between the years 1950-2010 [3], collectively this represents a significant loss of linguistic and cultural diversity. This position paper aims to explore the potential of AI-based language learning approaches that promote early exposure and appreciation of languages to ultimately contribute to the preservation of endangered languages, thereby addressing the urgent need to protect linguistic and cultural diversity.

## 1 INTRODUCTION

The preservation of endangered native languages is critical to maintaining cultural diversity, yet the number of languages in danger of disappearing is rapidly increasing. In this paper, we explore the potential of Artificial Intelligence (AI) in helping to preserve endangered native languages by promoting language appreciation among children. The paper argues that introducing children to their native language in a way that emphasizes appreciation is essential for language preservation efforts. By using AI-driven language learning tools, children can engage with their native language and culture in new and exciting ways, including the creation of bilingual storybooks, culturally specific illustrations, and virtual reality simulations that provide an immersive experience of the language and culture. The paper also highlights the importance of incorporating opportunities for children to co-create and co-design elements that go into the story, thereby empowering them to have agency in the language learning process alongside AI. Additionally, the paper examines the ethical considerations related to using AI in language learning, including accuracy, biases, and cultural appropriation. Ultimately, this position paper proposes AI-driven tools that leverage appreciation-based approach to language learning to promote language preservation and appreciation among children.

## 2 GENERATIVE AI

Preserving endangered native languages is crucial to maintaining cultural heritage, and it is especially important to introduce young children to their native language in a way that goes beyond basic communication and emphasizes language appreciation. For example, Artificial Intelligence (AI) has the potential to be utilized for the creation of bilingual storybooks that can be written in two languages, including the native language and a second language, such as English. In addition, generative AI like DALL-E or MidJourney can be used to develop illustrations that can highlight culturally unique expressions and aspects of the native language. Through an analysis of the story's text, AI can create vivid visual representations of the characters and settings that are inclusive of clothing, food, and decorations specific to the culture. By creating interactive images, the AI-generated illustrations can evoke inquisitiveness in children, prompting them to ask questions about the new elements they are seeing. Furthermore, the AI has the capacity to explain each aspect of the image while simultaneously weaving it into the storyline. By exposing children to the poetry, stories, fables, and other artistic expressions of their native language, they can develop a deeper understanding and appreciation for the phonology, grammar, and script of the language.

## 3 SOCIAL IMPLICATIONS

Utilizing AI-generated content that includes heritage and culturally specific elements can help promote a sense of pride and identity in their cultural heritage, which is particularly important for children who may be growing up in a multicultural environment. This can help foster a sense of connection and belonging, which can have positive implications for their overall well-being. It could facilitate connections between children from the same cultural background, promoting socialization and the formation of friendships based on shared cultural experiences. This could help foster a sense of community and belonging, which can be particularly important for children who may be living in areas where they are in the minority.

## 4 AI AND VIRTUAL REALITY

The utilization of AI has the potential to revolutionize the process of language learning for children by providing an interactive and personalized learning experience. AI-powered language learning tools can employ various media such as prompts, poetry, movies, comics, and other subtleties to enhance the child's engagement and retention of the language. For instance, virtual reality simulations have the potential to provide an immersive experience of the cultural context of the language. A virtual reality program can simulate a traditional market during a festival, offering children the opportunity to learn about different products, engage in haggling using the local language, and gain knowledge of cultural customs and traditions. Here, the AI algorithm can track the child's progress and provide adaptive exercises to reinforce the language skills needed to fully engage in the virtual reality experience. The AI can also provide contextual explanations of unfamiliar vocabulary or cultural references, which can enhance the child's understanding of the cultural context and help them to immersively participate in the experience. In addition to utilizing AI and virtual reality for language learning, it is important to incorporate opportunities for children to co-create and co-design elements that could go into a story. This approach empowers children to have agency in the language learning process alongside AI. For instance, children could collaborate with the AI-powered tool to create their own virtual reality environments that simulate real-life situations, encouraging the use of the target language in a more interactive and engaging way. This not only promotes language acquisition but also fosters creativity and problem-solving skills in children.

## 5 BRIDGING TRADITIONAL TEACHING METHODS AND AI

To optimize language learning outcomes, AI-centered design principles can leverage traditional language teaching methods, such as those found in older generation's school textbooks and authentic literature, to serve as a reference point for developing exercises that go beyond mere communication. The AI can be tailored to offer personalized recommendations to users based on their interests and interactions with the app's interactive elements, thereby bridging the gap between traditional language learning methods and modern technology. For instance, if a user engages with a poem and displays curiosity in understanding the meaning of each word, the app's AI algorithm can recognize the user's interest in the poem's use of personification to convey the beauty of nature, and subsequently recommend other poems that employ similar stylistic devices. This approach inspires the user to create their own poem using personification, thereby facilitating language learning and fostering creativity. Additionally, the app can suggest readings or exercises of prose and poetry based on user's interests and learning progress. By utilizing AI-centered design principles, the app not only provides personalized language learning experiences but also cultivates an appreciation for literature and poetry in the target language.

## 6 CHILD-AI COLLABORATION

AI-driven language learning tools have the potential to facilitate collaborative learning among children and elders by fostering teamwork to overcome language learning obstacles and engage in group-based activities such as language games and collaborative projects. For instance, an exercise could involve viewing the 2023 OSCAR Nominated documentary "The Elephant Whisperers" in its original language with friends/family, alongside the AI. The AI could respond in real time to the discussion, resolving uncertainties about the new cultural practices depicted in the documentary. Moreover, a generative AI could encourage the group of users to create a similar story by developing a digital storybook that explores the themes and cultural practices highlighted in the documentary and generating visually similar content for the storybook. This approach can aid language learning by reinforcing newly learned vocabulary and enhancing children's creativity while also promoting an understanding of AI as a knowledgeable persona that can guide in teaching and sharing new and interesting concepts. AI has come a long way in recent years, but it still requires training and data input to improve its language processing capabilities. In process of facilitating collaborative learning, AI can be trained to listen to folktales, poems, and songs told by elders to children, and develop training models based on cultural artifacts to better recognize and understand the nuances of the language. By employing AI to listen to and analyze these languages, it can serve as a valuable tool in language preservation efforts.

## 7 ETHICAL CONCERNS

However, there are concerns that need to be addressed, such as the ethical implications of using AI in language learning, ensuring the accuracy of the translations, and ensuring that the AI tools do not perpetuate harmful stereotypes or biases. These concerns can be addressed by involving native speakers and experts in the design and development process and regularly testing and refining the AI algorithms to improve their accuracy and effectiveness. In addition, since most AI like ChatGPT is a large language model, care must be taken to ensure that the AI does not necessarily generate new language that changes the historical languages. Rules and oversight should be included that may be able to frag when the AI is experimenting with new words and phrase types.

## 8 CONCLUSION

An appreciation-based approach can be a highly effective method for introducing young children to their native language, emphasizing the importance of not only learning the language itself but also understanding and appreciating the culture and values associated with it. This approach entails exploring the history, literature, and art associated with the language, as well as engaging with native speakers to gain insight into their cultural beliefs and practices. By promoting a deeper understanding and respect for the language and culture, this approach can ultimately enhance language proficiency and overall communication skills. With the aid of AI as the driving force behind language learning, this approach can be designed to function independently or in tandem with communicative language teaching and has the potential to save languages and cultures especially those that are in danger of being lost.

## ACKNOWLEDGEMENTS

I would like to extend my thanks to KidsTeam for the co-design research opportunities as well as their kind support and guidance that has been instrumental in shaping my research interests and academic growth.

## REFERENCES
[1] Languages | UNESCO WAL. (n.d.). En.wal.unesco.org. https://en.wal.unesco.org/languages
[2] UNESCO Institute for lifelong Learning. (2022, February 21). A decade to prevent the disappearance of 3,000 languages. UNESCO. https://www.iesalc.unesco.org/en/2022/02/21/a-decade-to-prevent-the-disappearance-of-3000-languages/
[3] Strochlic, N. (2018, April 16). Saving the World's Dying and Disappearing Languages. Culture; National Geographic. https://www.nationalgeographic.com/culture/article/saving-dying-disappearing-languages-wikitongues-culture